# On Test Sequence Generation using Multi-Objective Particle Swarm Optimization

Zain Iqbal [1], Kashif Zafar [1] Aden Iqbal [1] and Ayesha Khan [2]

[1] National University of Computer and Emerging Sciences, Lahore, Pakistan; (e-mail: zain.iqbal@nu.edu.pk; adeniqbal@gmail.com; kashif.zafar@nu.edu.pk)
[2] Forman Christian College (A Chartered University), Lahore, Pakistan; (e-mail: ayeshakhan@fccollege.edu.pk)

**ABSTRACT** Software testing is an important and essential part of the software development life cycle and accounts for almost one-third of system development costs [1]. In the software industry, testing costs can account for about 35% to 40% of the total cost of a software project [22]. Therefore, providing efficient ways to test software is critical to reduce cost, time, and effort. Black-box testing and White-box testing are two essential components of software testing. Black-box testing focuses on the software's functionality, while White-box testing examines its internal structure. These tests contribute significantly to ensuring program coverage, which remains one of the main goals of the software testing paradigm. One of the main problems in this area is the identification of appropriate paths for program coverage, which are referred to as test sequences. Creating an automated and effective test sequence is a challenging task in the software testing process. In the proposed methodology, the challenge of "test sequence generation" is considered a multi-objective optimization problem that includes the Oracle cost and the path, both of which are optimized in a symmetrical manner to achieve optimal software testing. Multi-Objective Particle Swarm Optimization (MOPSO) is used to represent the test sequences with the highest priority and the lowest Oracle cost as optimal. The performance of the implemented approach is compared with the Multi-Objective Firefly Algorithm (MOFA) for generating test sequences. The MOPSO-based solution outperforms the MOFA-based approach and simultaneously provides the optimal solution for both objectives.

## I. INTRODUCTION

Software testing remains the critical component of the software development life cycle (SDLC) and has a crucial impact on the success of a project [1]. There are numerous conventional and unconventional software problems, all of which can result in significant performance degradation if not addressed before software delivery. Such performance issues encompass faulty calculations, data manipulation, and data merging and reconciliation, which are further compounded by the fact that software testing is a time-consuming and costly task [2]. Estimates suggest that about 50% of the cost of software production is spent on the testing phase. At a very primary level, software testing can be divided into two broad categories: White-box testing [3] and Black-box testing [3]. The White-box approach focuses on the internal program structure of the software [4], whereas Black-box Testing views the software as a function that takes inputs and produces desired outputs [5]. Black-box testing, is based on requirements and functionality, encompassing all the components of a system and assessing what the software can achieve in terms of functionality also known as functional testing. It verifies the output for specific inputs without knowledge of the internal software structure.

On the other hand, white-box testing, also known as structural testing or glass-box testing, examines the logic and internal information within the software's code. It involves evaluating the coverage of code, coverage of branch, path coverage, and coverage of conditions to ensure the validity of the internal data structure. The main objective of this testing is to create test cases that achieve comprehensive program coverage. Optimal path testing combines branch coverage (testing all possible outputs of each branch in the program) and path testing (testing each selected path in the program at least once) [6]. Software testing encompasses three primary areas.

- Test data generation and test sequence generation, i.e., identification of suitable paths.
- Test execution of the software (System Under Test or SUT- using the generated test data).
- Evaluation of the test results obtained from the testing.

Three commonly used techniques for structural testing are control flow-based testing, data flow-based testing, and mutation testing [1], [2]. The software's control pathways are necessary to conduct control flow-based testing. Effective path selection becomes extremely critical since the total number of paths could be extensive even for testing a simple program. This phenomenon can lead to a very large number of test cases, hence significantly increasing testing costs [18]. Unified Modelling Language (UML) has already been employed for preliminary automation of the software testing process [18]; however, complete software coverage and path generation are still unaddressed research avenues. Various algorithms and techniques have been utilized to automate the generation of test data, including the Ant Colony Optimization (ACO) algorithm [14], Genetic Algorithm (GA) [32], and search-based test data generation [21]. These techniques aim to automatically generate test data to make software testing convenient in terms of cost and coverage while employing approaches such as swarm intelligence [21]. Other methods such as Cuckoo Search [33] and Artificial Bee Colony (ABC) [34] have also been utilized achieving good results for an automated and optimal test data generation. However, there exists certain limitations within these algorithms. Node redundancy remains a critical problem with these optimization techniques. Furthermore, as the size of the software increases, finding paths and test data gets more challenging since techniques such as ABC may become stuck in local search space and have a relatively high iteration count. On the other hand, there was limited consideration of path generation and coverage in the test data generation employing particle swarm optimization (PSO) [18]. Moreover, techniques such as ACO were used to try and build test paths [19], but the sequences produced demonstrated the problem of transition repetition which added tremendously to the overall cost of the software testing procedure.

Combinatorial testing (CT) [43] is an effective testing technique that allows for creating or designing an ideal or nearly optimal test suite with fewer test cases. The use of CT to identify defects brought on by interactions between configurations or components in a particular input space has proven to be successful [44]. According to empirical investigations on system defects, CT is, in certain ways, comparable to exhaustive testing [45]. Although CT has received much study and application, it does not always work well in certain circumstances, such as a system under test (SUT) whose test cases must be executed sequentially. For extensive testing, this SUT's testing specifications call for covering all factor interactions and all factor value sequences. According to the t-way combinatorial criterion, only the interactions between factor values can be effectively covered to a certain level when CT is used to create a test suite for this SUT; there is no adequate criterion for the value sequences of each factor [46].

In addition to the previously discussed challenges in software testing, selecting the optimal test sequence poses another significant task. To improve efficiency, effectiveness, and coverage while reducing time, cost, and effort, automating the testing process becomes crucial. This involves generating test cases that ensure comprehensive software coverage by identifying appropriate paths or test sequences. This paper presents the following contributions:

- A multi-objective optimization of test sequence generation with MOPSO and a comparison with other meta-heuristic algorithms
- Demonstrating that the proposed MOPSO technique fulfills the objective of attaining the highest priority with lowest cost for test sequences as compared to other SOTA algorithms

## II. ONLINE METHODS IN OPTIMAL TEST SEQUENCE GENERATION

Several research methodologies have explored the domain of meta-heuristic search techniques for optimal test sequence generation [7]. The existing works in the literature have utilized various methods and fitness functions for optimal test sequence generation in the field of Software Engineering (SE) and Artificial Intelligence (AI). To serve the purpose of protocol specification, an Extended Finite State Machine (EFSM) was proposed for generating test sequences [8]. The methodology introduced an approach referred to as Unified Test Sequence (UTS) to test protocols with the EFSM and generate a single test sequence. The objective was to obtain a short test sequence. However, a shortcoming of the methodology was the redundancy of the generated sequence. Busser et al. [10] proposed automatic test sequence generation by introducing a new method called Test Sequence Vectors (TSVs) for specification-based test generation. However, in an important discussion of this method, it was noted that test vector generation is more convenient in creating test sequences but does not provide the user with a complete system scenario [9]. Moreover, a method for generating test sequences from the description of LUSTRE was proposed, and the GATEL tool was employed to automate test sequence generation [11]. Another study on test sequence generation [13] introduced a novel approach for generating automated test sequences using a formal requirements model. Comprehensive test cases were generated using the model checker. However, the state space explosion was not considered despite it being a significant problem in model checking of software artifacts. Jervan et al. [12] investigated the feasibility of high-level fault models for generating test sequences. The goal was to facilitate software engineers in the early stages of the design phase to estimate circuits in terms of their testability and test sequences. The methodology utilized Automatic Test Path Generation (ATPG) algorithm to evaluate the effectiveness of high-level test generation. The generated test sequences were found to be efficient in sequence generation, using techniques that employed constraint-solving algorithms.

For optimum test sequence generation, a meta-heuristic approach, "Ant Colony Optimization (ACO)", was proposed for state-based software [14], [19]. Test sequences were automatically generated using ACO, while UML state diagrams were employed to define test adequacy standards. However, the methodology could not compensate for the redundancy in the paths [16], [17], [18]. Another study by Srivastava et al. [15] proposed an optimized test sequence generation with ACO from usage models that visualized software in terms of states during model-based testing. However, redundancy within the test sequences was not considered here either. A multi-objective evolutionary algorithm, M-GEOvsl, was proposed and used to generate feasible test sequences based on Pareto optimality [20]. However, the flaw in the methodology was that there was no proper mechanism to deal with the situation where no slice could be found for the model, and a redundant path was generated. In addition, the Firefly algorithm (FA) [21] was presented as a research approach to generate optimal test sequences. The FA algorithm was enhanced by incorporating an objective function, cyclomatic complexity, and a random function. Initially, it was employed to generate an set of optimal test sequences by using a state transition diagram, which were subsequently prioritized based on the average brightness values. This technique achieved the code coverage through optimal and non-redudant paths and exhibited superior performance compared to the previous ACO approach. However, a limitation of this method was that it addressed a unidirectional optimization problem. In a study by Windisch et al. [22], Particle Swarm Optimization (PSO) was applied during the software testing process with comparative analysis using the Genetic Algorithm (GA). It was concluded that PSO was less expensive in generating test sequences and outperformed GA. In the research technique presented by Li et al. [25], PSO was deployed to automatically generate test data for all paths to reduce the overall cost of software testing. The results exhibited better performance of PSO in terms of path generation as compared to the previous techniques. A variant of the Firefly algorithm known as the Multi-Objective Firefly Algorithm (MOFA) has been introduced to tackle optimization problems. [23]. MOFA was developed to solve multi-objective optimization problems and to further find Pareto-optimal and non-dominated solutions, especially for better optimization of multiple objects.

Coello et al. [24] proposed MOPSO, an extension of the PSO algorithm that can be used to solve multi-objective optimization problems. Iqbal et al. [28] proposed a technique for multi-criteria optimization of test sequence generation using the Multi-Objective Firefly Algorithm (MOFA). In the proposed work, the optimal test sequence was generated while considering the objectives of Path Priority and Oracle Cost simultaneously. Sharma et al. [29] generated an optimal test sequence by employing a swarm intelligence method called River Formation Dynamics (RFD). The proposed technique generated test sequences in the control flow graph of the program, which were evaluated as the software testing progressed. The method was based on various RFD parameters and constants and provided a complete path coverage without edge or transition redundancy. However, the method failed when considering multiple parameters simultaneously and proved incapable of considering the problem as a multi-objective task. Saha et al. [30] generated test sequences using a nature-inspired meta-heuristic, the Moth Flame Optimization Algorithm (MFO). The main goal was to provide fewer and unique test cases to increase optimization. MFO proved suitable for large object-oriented software applications but did not provide promising results for smaller applications. Furthermore, the technique was not tested on multi-objective problems and data representing a combination of small and large software solutions.

Multi-objective evolutionary algorithms (MOEAs) have emerged as one of the most effective strategies for solving such MOPs since they can approximate many non-dominated solutions in a single run [37]. The multi-objective evolutionary algorithm based on decomposition (MOEA/D) is one of the most modern and successful approaches [38]. An MOP is divided into several smaller scalar optimization problems in MOEA/D, and these smaller problems are handled simultaneously. Due to the preservation of subproblem variety and knowledge sharing among individuals of a neighbourhood, MOEA/D may obtain evenly dispersed solutions along a Pareto front (PF). Additionally, the efficiency of MOEA/D is demonstrated through real-world applications like route planning [39], economic emission dispatch [39], and wireless sensor networks [39]. However, recent research in [40, 41, 42] revealed that MOEA/D only successfully solves MOPs with simple PFs and struggles when faced with MOPs with complex PFs that display characteristics like a long tail or a sharp peak. Different versions of MOEA/D have been developed in recent years using various decomposition techniques and evolutionary algorithms, including MOEA/D-DE [37], MOEA/D-DRA [38], MOEA/D-XBS [37], and MOEA/D-GR [41]. There is no one MOEA/D version that combines the many unique advantages of the many versions, although different MOEA/D variants are available in the literature.

The CT-based testing is sequence-based t-way testing for SUTs with n input events, where each event occurs exactly once throughout a test. Kuhn et al. was the first to employ sequence-based t-way testing. He published

Sequence Covering Arrays (SCAs) in accordance with covering arrays [47] preceded by a "quick and dirty" (QnD) [48] technique, which is based on a greedy algorithm and has the potential to develop. He then put forth a modified greedy method to manage the limitations between event pairs [49] while further testing labelled transition systems using SCAs [50]. He provided several SCA generation algorithms using the suggested t-way sequence coverage criterion. These algorithms mark the first attempt to methodically investigate potential courses of action for resolving the issue of t-way test sequence generation in a broad sense. As an extension of current t-way strategies, Zamli et al. examined the sequence-based fixed and variable strength testing and concluded that there was unquestionably room for improvement, notably for the t-way sequence coverage testing [51]. Afterwards, [52] introduced a sequence-based t-way interaction testing technique employing the bees algorithm. When tested against QnD, the suggested algorithm yields encouraging results. [53] introduced the t-way Event-Driven Input Sequence Test Case (EDISTC-SA) generator, a technique for creating a t-way event-driven test suite based on simulated annealing. Farchi et al. established the ordered constraints and interaction coverage requirements [46] and described a test as an ordered tuple of input parameter values. Then, a useful approach was put forth and tested on various real-world systems for creating test suites with minimal sizes that satisfy the ordered interaction coverage criteria. SCAs have valuable applications in testing as they can lower the cost of testing by reducing the number of tests since exhaustive testing always involves an enormous number of tests in comparison.

Nguyen et al. [54] outlined a novel strategy combining model-based and combinatorial testing to create executable and practical test sequences. The method begins with a finite state model and creates executable routes describing events that should be performed on the SUT. The equivalence classes of a classification tree are then created from these paths [55]. Finally, the events are represented by the first child classifications of the tree's root node, and each classification's classes are the optional values of the corresponding event. The classification tree is then utilized to build executable test cases using t-way testing for each executable path. Based on equivalence partitioning, the classification tree method [55] and a model-based black-box test design tool [56] were proposed. It is frequently employed for systematic test design and test case description. TESTONA [57] and TESSY [58,59] are mature products that use the classification tree method to construct test sequences. The input domain of a SUT is classified using the classification tree approach under several parameters that the tester has determined to be equivalent. Disjoint and complete classifications are divided for each aspect. The following classification of each aspect is represented graphically using a tree-like structure [55]. Recursive classification can be applied to categorize classes resulting from these classifications [55], as they are disjointed abstractions of particular input levels for testing reasons [60]. A test suite's tree serves as the combination table's head, while the table's test cases serve as its body. Test cases, which correspond to the testing task's test phases, are created by merging classes from several classifications. Test cases are typically carried out in order during the test run. In terms of functional requirements, the testing design of the classification tree technique has been utilised extensively for embedded systems [61], embedded automotive systems [61,56], and online applications [62]. Additionally, "Modbat" and Microsoft's "Spec Explorer" are tools for creating model-based test cases. State transitions coverage test cases can be generated by "Modbat" [63]. By using traversal techniques to produce automated test cases, Microsoft's "Spec Explorer" can accomplish a type of transition coverage and allow testers to identify requirements violations with the least amount of manual work [64].

Despite extensive research in this area, the effective generation of test sequences remains a domain with great potential for improvement, primarily to soundly address redundancy and multiple optimization problems. This paper presents a novel approach that employs a Multi-Objective Particle Swarm Optimization algorithm to generate optimal and non-redundant test sequences efficiently.

### III. MATERIALS AND METHODS

The meta-heuristic algorithms ACO and GA do not perform well when applied to the test sequence generation problem due to redundancy issues. PSO, in particular, showed competitive performance when applied to the test path generation problem compared to GA. It also has a robust local search capability, fast convergence, and lower computational cost. The optimal test sequence generation problem can be viewed as a multi-objective optimization problem with two symmetrical and distinct objectives that are non-dominated for the given problem..

- Applying Multi-Objective Particle Swarm Optimization (MOPSO) on test sequence generation considering multiple constraints [21] [27]:

  Path Priority: $F1(x) = [100 / (CC_i * rand())]$

  Oracle Cost: $F2(x) = [tc / (Bp * MaxPriority(P, A))]$

- Perform comparative study of MOPSO with (MOFA) on the problem of test sequence generation.

*A. CONSTRAINTS*

The two objectives, Path Priority and Oracle Cost, considered for optimization are conflicting since a lower priority corresponds to a lower cost. The goal is to find a solution simultaneously with the highest priority and the lowest cost. Pareto optimization is suitable for problems with multiple competing constraints because it can find a balance within them [26].

1) Path Priority: Test sequences covering the non-redundant paths while holding the highest mean brightness value and providing the complete coverage of the state transition diagram (STD).

2) Oracle Cost: Execution of test sequence reflects the expense of verifying the system's behavior through those specific sequences and the cost associated with those sequence are called oracle cost. In software systems, testing costs pose a considerable performance challenge, and therefore, the objective is to minimize them.

*OBJECTIVE FUNCTION*

The objective function determines how effectively a program can solve a problem, and based on this function, appropriate solutions are passed on to the next generation of solutions. The objective function would comprise following equations [21][27]:

$$F(x) = \{F1(x), F2(x)\}$$

Where F1(x) indicates the Objective Path Priority given as,

$$F1(x) = [100 / (CC_i * rand())]$$

Where i represent the node, CC= no. of decision nodes + 1, rand= $[[N_i – i] – 0.1]$, N= Total number of nodes

Furthermore, F2(x) indicates the Oracle Cost objective,

$$F2(x) = [tc / (B_p * MaxPriority(P, A))]$$

Where, the value of tc indicates the number of test sequences generated, Bp value indicates the number of edges/branches, MaxPriority (P, A) indicates the maximum priority achieved by the algorithm A in the program P. The proposed algorithm is presented in the Figure 1.

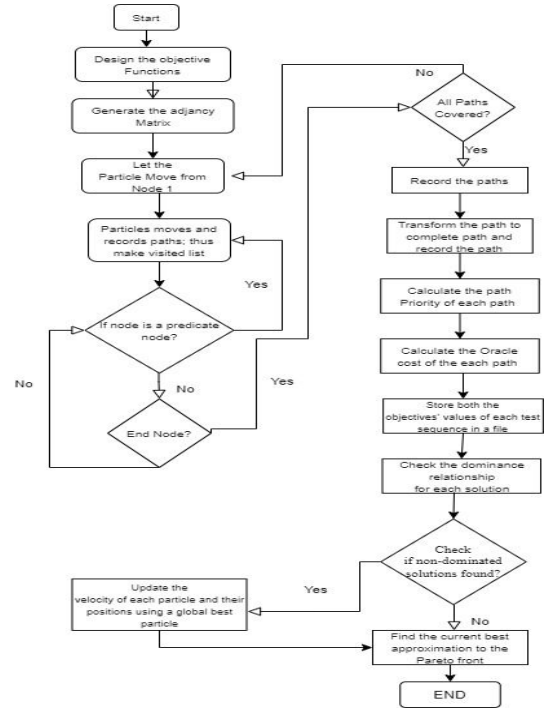

**FIGURE 1.** Algorithm for calculating the objective function of the MOPSO. Path Priority and Oracle Cost values are calculated to optimize the function with the objective of achieving high priority and low cost simultaneously.

### B. MULTI-OBJECTIVE OPTIMIZATION OF TEST SEQUENCE GENERARION WITH MOPSO

To perform multi-objective test sequence optimization with MOPSO. The utilization of the state transition diagram was employed to evaluate the Path Priority and Oracle Cost for the generated test patterns. The algorithm for multicriteria optimization using MOPSO is presented below for illustration. Each node represents a state of the program being tested for software evaluation, while the edges, denoting the transitions between states, represent the paths to be evaluated with PSO. Adjacency matrices are generated to determine the uniqueness of the paths generated from the state transition diagrams, which is an important metric for determining the effectiveness of the paths for software testing. With PSO, different particle numbers are pushed through the edges between nodes in the diagram, recording the nodes visited for test sequencing. After traversing all possible paths, objective function values are calculated for each path, i.e., Path Priority and Oracle Cost. The non-dominated solutions among the generated sequences are also determined to provide an optimum solution.

**Algorithm 1: Multi-Objective Optimization of Test Sequence Generation using (MOPSO)**

**Input:** State Transition Diagram
**Output:** Test Sequences

1. Designing the objective function
2. Evaluate Cyclometric Complexity of each node
3. Generation of the Guidance and Adjacency Matrix
4. **for** Particle move from node 1 **do**
5. Particle make visited list of the visited paths and save it
6. *If* node is the predicate node? **Then**
7.    go to step 16
8. *else*
9.    go to step 11
10. *end if*
11. *If* node is end node? **Then**
12.    go to step 17
13. *else*
14.    go to step 5
15. *end if*
16. Use the value of guidance matrix to move particle to the next node
17. *If* all the paths are covered? **Then**
18.    go to step 22
19. *else*
20.    go to step 4
21. *end if*
22. Recode the paths
23. Calculate the path priority by calculating the mean brightness value of each path
24. Calculate the value of oracle cost for each path
25. Save the values for both the objectives' of each test sequence in a file
26. For each solution check the dominance relationship
27. *If* non-dominated solutions found? **Then**
28.    go to step 34
29. *else*
30.    go to step 33
31. *end if*
32. Update the velocity of each particle and their positions using a global best particle
33. Find the current best approximation value to the Pareto front
34. *end for*

### D. MULTI-OBJECTIVE OPTIMIZATION OF TEST SEQUENCE GENERATION WITH MOFA [29]

A state transition diagram was deployed for multi-objective test sequence optimization with MOFA to determine the Path Priority and Oracle Cost for the generated test patterns. The algorithm for multi-objective optimization with MOFA is explained in more detail below. A similar setup to PSO was utilized for MOFA optimization, and objective function computation, where the states act as nodes and the transitions are represented by the edges between the nodes. Different firefly numbers were rendered through the edges between the nodes in the diagram. The visited nodes were tracked in a visited list to illustrate the different test paths for optimal sequence generation. After all potential paths were explored, objective function values were calculated for each path, i.e., Path Priority and Oracle Cost, and the non-dominated solutions were ascertained among the generated sequences.

**Algorithm 2: Multi-Objective Optimization of Test Sequence Generation using (MOFA) [29]**

**Input:** State Transition Diagram
**Output:** Test Sequences

1. Designing the objective function
2. Evaluate Cyclometric Complexity of each node
3. Generation of the Guidance and Adjacency Matrix
4. **for** Firefly move from node 1 **do**
5. Firefly make visited list of the visited paths and save it
6. *If* node is the predicate node? **Then**
7.    go to step 16
8. *else*
9.    go to step 11
10. *end if*
11. *If* node is end node? **Then**
12.    go to step 17
13. *else*
14.    go to step 5
15. *end if*
16. Use the value of guidance matrix to move firefly to the next node
17. *If* all the paths are covered? **Then**
18.    go to step 22
19. *else*
20.    go to step 4
21. *end if*
22. Recode the paths
23. Calculate the path priority by calculating the mean brightness value of each path
24. Calculate the value of oracle cost for each path
25. Save the values for both the objectives' of each test sequence in a file
26. For each solution check the dominance relationship
27. *If* non-dominated solutions found? **Then**
28.    go to step 34
29. *else*
30.    go to step 33
31. *end if*
32. Generate random weights
33. Find the current best approximation value to the Pareto front
34. *end for*

## IV. EXPERIMENTATION AND RESULTS

Experiments were performed on the ATM-one Transaction dataset and its respective State Transition Diagram [21]. Multi-Objective Particle Swarm Optimization (MOPSO) was used to generate test sequences while considering both single and multi-objective evaluation metrics [21][27].:

- Path priority: $F1(x) = [100/ (CC_i * rand ())]$

- Oracle Cost: $F2(x) = [tc / (Bp * MaxPriority (P, A))]$

| ATM-One Transaction State Transition Diagram | |
|---|---|
| Number of nodes | 8 |
| Number of edges | 13 |
| Number of the predicate nodes | 4 |
| Number of the test sequences generated | 6 |
| Particles and Fireflies deployed | 3,5,7,10,15,20 |

A comparative study was conducted between MOPSO

**TABLE 1. State transition diagram details for the ATM transaction use case.** Nodes represent the states, while the edges represent the transitions between the connected nodes. Test sequences are generated with the transition paths, while ensuring that there is no redundancy. Both single and multi-objective optimization algorithms utilize particles and fireflies to generate test sequences.

and other meta-heuristic algorithms for the test sequence generation problem. For the comparative study of bio-inspired algorithms, FA and PSO were employed. Both techniques exhibited a better convergence and consistent training mechanism. The sequences generated using Particle Swarm Optimization (PSO), Multi-Objective Particle Swarm Optimization (MOPSO), Firefly Algorithm (FA) and Multi-objective Firefly Algorithm (MOFA) were similar. Test sequences were also created for a single target using Particle Swarm Optimization and the Firefly Algorithm. The objective of the "Path Priority" in generating these test sequences was to achieve complete coverage of the STD and assign the maximum mean brightness value or priority, representing its significance or criticality among the generated optimal test sequences. The test sequences produced by these multicriteria techniques consisted of non-redundant paths. To conduct a comparative analysis of the optimization algorithms, varying numbers of particles and fireflies were utilized in the experiments. For the implementation of multi-objective optimization, optimal test sequences were generated considering two objectives: Path Priority and Oracle Cost, using the Multi-Objective Particle Swarm Optimization and Multi-Objective Firefly Algorithm. Additionally, the algorithms discovered test sequences that were not dominated by other solutions, known as non-dominated solutions are shown by bold markers in the diagrams and tables in the discussion of the experiments. The algorithm and state diagram of the ATM-One

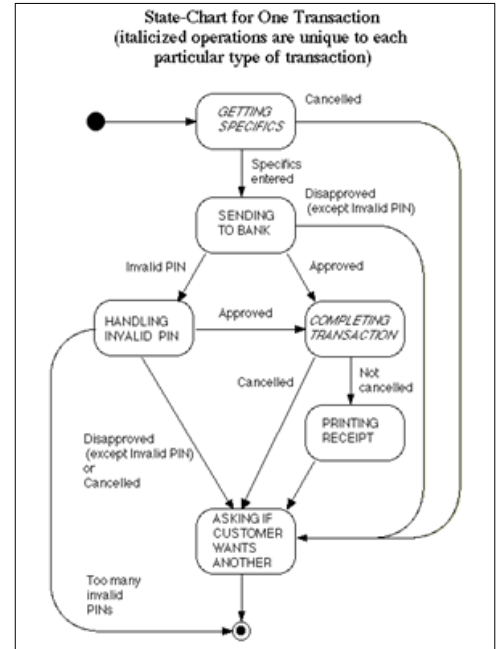

Transaction dataset are exhibited in Figure 2. The details of the state diagram are exhibited in the Table 1.

**FIGURE 2 State Transition chart for ATM Transaction dataset.** Different states are utilized during an atm transaction, encompassing sending of the transaction specifics, receival and receipt of the transaction amount, and user request for another transaction. These states form the basis of the software testing procedure for the atm transaction system employed for the study

For comparative analysis, two methods [21] and [28] were implemented in the experimental setup; Particle Swarm Optimization and Firefly Algorithm. The optimal test sequence was created in MATLAB using the four variants of the two main algorithms: Single Objective Particle Swarm Optimization, Multi-Objective Particle Swarm Optimization, Single Objective Firefly Algorithm, and Multi-Objective Firefly Algorithm. Multiple runs were performed, e.g., five runs for each particle and firefly in the experiment to test the techniques with a range of particles and fireflies for comparative analysis.

### A. IMPLEMENTATION OF PARTICLE SWARM OPTIMIZATION WITH SINGLE OBJECTIVE

The test sequences were generated utilizing particle swarm optimization, with the primary objective of achieving the complete coverage of the STD and obtaining the maximum mean brightness value or attaining priority among the generated optimal test sequences to signify their significance or criticality. These generated test sequences comprised non-redundant paths. For experimental purposes, a different number of particles were used in each run with the same data set. Table 2 shows the results of particle swarm optimization on ATM-One Transaction State Transition Diagram. The average values in Table 2 show that using multiple particles on the dataset resulted in higher path priority values for generating test sequences. Furthermore, Figure 3 also shows the Objective values (Path Priority) for test sequence generation. The main goal was to obtain values that cover the state transition diagram entirely and have the highest mean global value (or priority), achieved at particle number 20.

| Optimal Test Sequences Generated | Independent paths generated | Path Priority | | | | | |
|---|---|---|---|---|---|---|---|
| | | P = 3 | P = 5 | P = 7 | P = 10 | P = 15 | P = 20 |
| 1,2,7,8 | 1,2,7,8 | 0.100 | 0.150 | 0.237 | 0.362 | 0.550 | 0.737 |
| 2,3,7 | 1,2,3,7,8 | 0.100 | 0.170 | 0.260 | 0.390 | 0.590 | 0.780 |
| 3,5,7 | 1,2,3,5,7 | 0.127 | 0.193 | 0.290 | 0.431 | 0.637 | 0.856 |
| 3,4,7 | 1,2,3,4,7 | 0.100 | 0.187 | 0.285 | 0.425 | 0.627 | 0.850 |
| 3,5,6,7 | 1,2,3,5,6,7 | 0.100 | 0.185 | 0.278 | 0.414 | 0.621 | 0.835 |
| 4,8 | 1,2,4,8 | 0.087 | 0.150 | 0.250 | 0.362 | 0.550 | 0.720 |
| **Average Value** | | **0.102** | **0.172** | **0.267** | **0.397** | **0.596** | **0.797** |

**TABLE 2 PSO generated test sequences extracted from the ATM state transition diagram.** Test sequences were generated with the single objective, maximizing path priority value while providing full coverage of the state transitions. Average path priority values are recorded for a range of particle numbers, which establishes the increasing trend of the path priorities with the increase in particle numbers.

### B. IMPLEMENTATION OF MULTI-OBJECTIVE PARTICLE SWARM OPTIMIZATION WITH MULTIPLE OBJECTIVES

Optimal test sequences were generated using Multi-Objective Particle Swarm Optimization considering two objectives: Path Priority and Oracle Cost. In addition, test sequences that do not dominate other solutions were also found within the state diagram. Non-dominated solutions and are represented by bold markers in the graphs and tables. As can be observed from Table 3, different numbers of particles were used in each run on the same data set for experimental purposes. Tables 3 and 4 exhibits the results of Multi-Objective Particle Swarm Optimization on ATM-One Transaction State Transition Diagram, where the last row displays the average cost value. In addition, Table 6 displays the values for all non-dominated solutions of each objective, i.e., the average of all non-dominated solutions or test sequences, each shown in bold. Furthermore, Figure 4 demonstrates the objective values, including Oracle Cost and Path Priority, as well as the average values for all non-dominated optimal test sequences.

As demonstrated in Figure 4, the path priority values increase with an increase in the particle number. However, the cost for the path fluctuates, with the optimum balance, demonstrated with a high priority and a low cost, achieved on a particle number of 15. With MOPSO, the highest cost was incurred by a particle

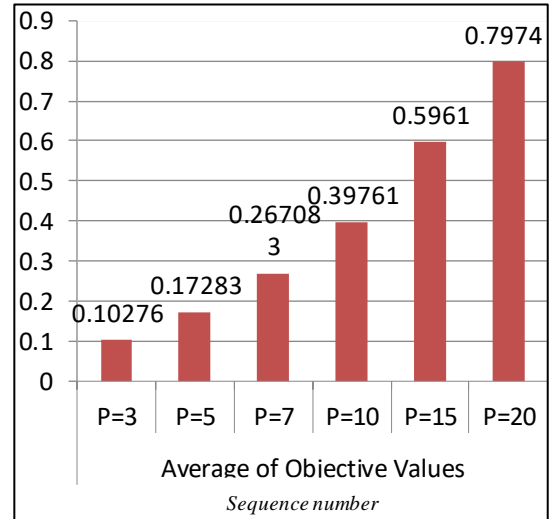

number of 3, whereas the highest priority was achieved on a particle number of 20.

**FIGURE 3 ATM-One Transaction STD.** Path Priority average values for all particles deployed using PSO. A single objective of priority path was used for generating test sequences with different particle numbers. Path Priority demonstrated an increase with the increase in particle number. The highest average path priority was achieved at a particle number of 20 with PSO.

| P=3 | | P=5 | | P=7 | |
|---|---|---|---|---|---|
| Path Priority | Cost | Path Priority | Cost | Path Priority | Cost |
| **0.10010** | **2.44600** | **0.15000** | **0.41580** | **0.23750** | **0.86940** |
| **0.10030** | **0.19320** | **0.17000** | **0.59760** | **0.26010** | **0.88690** |
| 0.12730 | 2.80260 | 0.19380 | 2.07430 | 0.29050 | 2.04060 |
| **0.10050** | **2.22400** | **0.18750** | **0.41120** | **0.28570** | **0.37140** |
| **0.10090** | **0.58930** | **0.18570** | **0.24190** | **0.27860** | **1.75000** |
| 0.08750 | 2.86680 | 0.15001 | 2.21280 | 0.25010 | 1.26600 |
| 0.10582 | 1.65102 | 0.16857 | 0.41843 | 0.26668 | 1.38673 |

**TABLE 1 Objective values of PSO generated sequences for particles numbers of 3, 5, and 7.** Bold values denote the objective values of non-dominated solutions that are not dominated or impacted by other test sequences generated by MOPSO. The last row highlights the average cost for the non-dominated solutions since they non-redundant, thus being applicable for a comprehensive and reliable software testing procedure.

| P=10 | | P=15 | | P=20 | |
|---|---|---|---|---|---|
| Path Priority | Cost | Path Priority | Cost | Path Priority | Cost |
| 0.36250 | 1.06970 | **0.55010** | **0.61500** | 0.73750 | 0.60660 |
| **0.39010** | **0.25200** | **0.59010** | **0.33920** | **0.78000** | **0.32290** |
| 0.43130 | 1.28200 | **0.63750** | **0.18530** | 0.85620 | 0.48850 |
| **0.42500** | **0.22470** | **0.62750** | **0.21460** | 0.85000 | 0.61810 |
| 0.41430 | 1.29770 | **0.62140** | **0.05610** | 0.83570 | 0.68330 |
| **0.36250** | **0.35180** | 0.55000 | 0.81380 | **0.72500** | **0.59940** |
| 0.39253 | 0.59147 | 0.60532 | 0.28204 | 0.78707 | 0.47027 |

**TABLE 2 Objective values of PSO generated sequences for particles numbers of 10, 15, and 20.** Bold values denote the objective of non-dominated solutions that are not dominated or impacted by other test sequences generated by MOPSO. The last row highlights the average cost for the non-dominated solutions since they non-redundant, thus being applicable for a comprehensive and reliable software testing procedure.

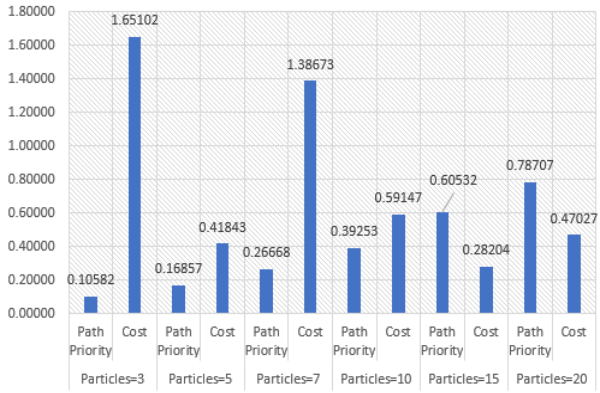

**FIGURE 4 ATM-One Transaction STD with MOPSO.** Objective and average values are demonstrated for all particles deployed using MOPSO. A multi-objective parameter is set, which considers both the Path Priority and Oracle Cost. Test with 20 particles achieved the best path priority. Moreover, the greatest cost was incurred by a particle number of 3 on MOPSO. Most importantly, the optimum balance between the path priority and cost was achieved by particle value of 15.

| Optimal Test Sequences Generated | Independent paths generated | Objective | |
|---|---|---|---|
| | | Path Priority | Cost |
| 1,2,7,8 | 1,2,7,8 | 0.3625 | 1.0697 |
| **2,3,7** | **1,2,3,7,8** | **0.3901** | **0.2520** |
| 3,5,7 | 1,2,3,5,7 | 0.4313 | 1.2820 |
| **3,4,7** | **1,2,3,4,7** | **0.4250** | **0.2247** |
| 3,5,6,7 | 1,2,3,5,6,7 | 0.4143 | 1.2977 |
| 4,8 | 1,2,4,8 | 0.3625 | 0.3518 |
| Average Values | | 0.39253 | 0.59146 |

**TABLE 3 ATM-One Transaction STD: Objective Values with Particles = 10 using MOPSO.** Objective values and average values are demonstrated for all the particles deployed using MOPSO. Average objective values calculated are displayed for all the optimum paths generated from the state transition diagram for a particle number of 10. Paths in bold are the non-dominated solutions which are not dominated by any other test sequences.

For experimental purposes, an optimal test sequence was created using a varying number of particle runs. The results showing the objective values with MOPSO are exhibited in Table 5. Also, an important observation was made in Figure 5, which exhibits an inverse relationship between the highest path priority and the cost incurred by MOPSO at a particle count of 10. The highest path priority had higher costs, which deviates from the goal of achieving lower costs for high path priorities to manifest an effective software testing process.

### C. FIREFLY ALGORITHM IMPLEMENTATION WITH SINGLE OBJECTIVE

For the following experiment, test sequences were generated using the Firefly algorithm. The algorithm had only the single objective, which was to generate test sequences that achieved complete coverage of STD and possessed the highest mean brightness value or priority, indicating their importance or criticality between the generated optimal test sequences. The generated test sequences from the STD were non-redundant paths. For the experiments, a different number of fireflies were used in each run with the same dataset. Table 7 exhibits the results obtained with the Firefly algorithm on ATM-One Transaction State Transition Diagram.

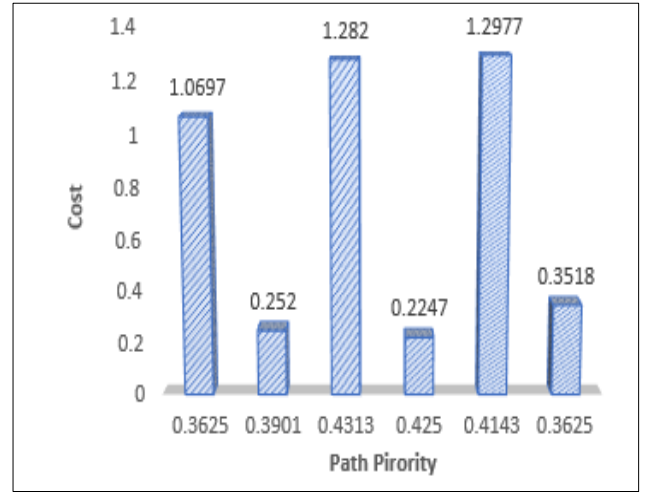

**FIGURE 5 Graphical representation of optimal test sequence generated using a particle number of 10.** As demonstrated, there exists a non-ideal relationship between the highest cost and its path priority for a particle number of 10 while employing MOPSO for generation optimum test sequences. The highest priority path exhibited the highest cost which is not an optimum condition for software testing.

Another important observation can be solicited from Figure 5, which exhibits an inverse relationship between the highest path priority and the cost of MOPSO at a particle count of 10. The highest path priority had higher costs, deviating from the goal of achieving lower costs for high path priorities.

### C. FIREFLY ALGORITHM IMPLEMENTATION FOR SEQUENCE GENERATION WITH SINGLE OBJECTIVE

For the following experiment, test sequences were generated by the Firefly algorithm. The algorithm focused on a single objective known as "Path Priority" to generate test sequences. These sequences aimed to achieve complete coverage of the state transition diagram while having the highest mean brightness value or priority, signifying their significance or cruciality among the optimal test sequences generated. The resulting sequences consisted of non-redundant paths and included non-dominant solutions, which were not overshadowed by other sequences produced by the Firefly algorithm.

| Optimal Test Sequences Generated | Independent paths generated | Path Priority | | | | | |
|---|---|---|---|---|---|---|---|
| | | FF=3 | FF=5 | FF=7 | FF=10 | FF=15 | FF=20 |
| 1,2,7,8 | 1,2,7,8 | 21.25 | 20.25 | 19.89 | 19.60 | 19.38 | 19.27 |
| 2,3,7 | 1,2,3,7,8 | 18.20 | 17.44 | 17.14 | 16.90 | 16.73 | 16.64 |
| 3,5,7 | 1,2,3,5,7 | 8.60 | 8.48 | 8.40 | 8.36 | 8.33 | 8.32 |
| 3,4,7 | 1,2,3,4,7 | 4.86 | 4.84 | 4.80 | 4.78 | 4.77 | 4.77 |
| 3,5,6,7 | 1,2,3,5,6,7 | 14.38 | 14.10 | 13.97 | 13.86 | 13.80 | 13.75 |
| 4,8 | 1,2,4,8 | 4.50 | 4.50 | 4.46 | 4.45 | 4.43 | 4.43 |
| Average Value | | 11.96 | 11.60 | 11.44 | 11.32 | 11.24 | 11.20 |

**TABLE 4 Firefly algorithm test sequences extracted from the ATM state transition diagram.** Test sequences were generated with the single objective of maximizing path priority value while providing full coverage of the state transitions. Average path priority values are displayed for a range of particle numbers, which establishes a nominal

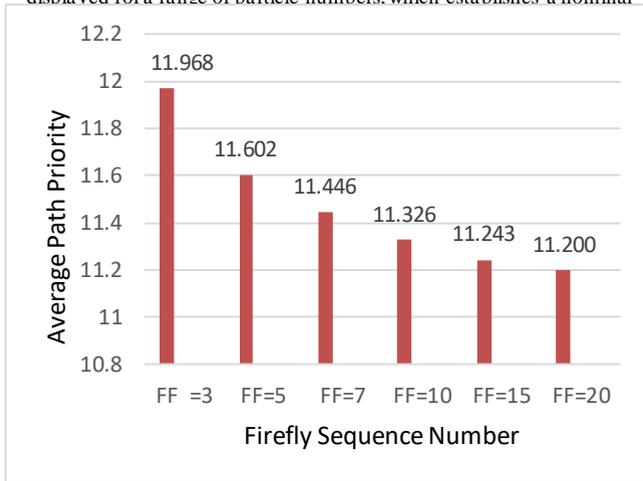

**FIGURE 6 ATM-One Transaction STD For Firefly algorithm.** Average of Path Priority values are displayed for all test sequences deployed using FF. A multi objective of Priority Path and Oracle Cost was used for generating test sequences. The highest priority was achieved at an FF number of 3.

For the experiments, different numbers of fireflies were used in each run with the same dataset. Table 7 exhibits the results obtained with the Firefly algorithm on ATM-One Transaction State Transition Diagram. The average values in Table 6 indicate that by using multiple Fireflies on the dataset, higher path priority values for generating test sequences can be achieved. Furthermore, Figure 6 represents the Objective values (Path Priority) for test sequence generation, showing the highest Path Priority at firefly number 3.

### D. IMPLEMENTATION OF MULTI-OBJECTIVE FIREFLY ALGORITHM WITH MULTIPLE OBJECTIVES

Optimal test sequences were generated using the Multi-Objective Firefly algorithm considering the multiple objectives of Path Priority and Oracle Cost. Test sequences were found that did not dominate other solutions. These solutions are referred to as non-dominated solutions and are represented by bold markers in the graphs and tables. For the experiments, a different number of Fireflies were used in each run with the same dataset. Table 8 and Table 9 show the results of the Multi-Objective Firefly algorithm on the ATM-One Transaction State Transition Diagram. The average values were calculated for all non-dominated solutions of each objective, which is the average of all bold values. In addition, Table 10 shows the values for all non-dominated solutions of each objective with the average of all bold, non-dominated solutions.

| Optimal Test Sequences Generated | Independent paths generated | Path Priority | | | | | |
|---|---|---|---|---|---|---|---|
| | | FF=3 | FF=5 | FF=7 | FF=10 | FF=15 | FF=20 |
| 1,2,7,8 | 1,2,7,8 | 21.25 | 20.25 | 19.893 | 19.601 | 19.383 | 19.275 |
| 2,3,7 | 1,2,3,7,8 | 18.201 | 17.44 | 17.143 | 16.901 | 16.733 | 16.640 |
| 3,5,7 | 1,2,3,5,7 | 8.600 | 8.480 | 8.401 | 8.36 | 8.333 | 8.321 |
| 3,4,7 | 1,2,3,4,7 | 4.867 | 4.840 | 4.801 | 4.780 | 4.773 | 4.771 |
| 3,5,6,7 | 1,2,3,5,6,7 | 14.389 | 14.101 | 13.976 | 13.867 | 13.801 | 13.758 |
| 4,8 | 1,2,4,8 | 4.501 | 4.501 | 4.464 | 4.450 | 4.4331 | 4.438 |
| Average Value | | 11.9676 | 11.6016 | 11.4460 | 11.3261 | 11.2425 | 11.2001 |

**TABLE 7 ATM-One Transaction Objective and average values with 3, 5, and 7 fireflies deployed using MOFA.** Non-dominated values are highlighted in bold. Test sequences are evaluated on a multi-objective condition, encompassing both Path Priority value and cost. Best set of values were attained on a firefly value of 5, demonstrating the optimum conditionality of high path priority and low cost for test sequence generation.

The MOFA experiments provide objective values for both Path Priority and Oracle Cost. Figure 7 exhibits the MOFA objective values that include both Oracle Cost and Path Priority, and also displays the average values for all non-dominated optimal test sequences for all particle values used. Various Firefly values were utilized to perform a comparative study of the MOFA algorithm with the goal of producing an optimal test sequence with the highest path priority and lowest cost. FF Number 5 showed the best balance between the two objectives, with the highest path priority of 17.26 and the lowest possible cost of all other FF values at 10.02. In addition, the optimal test sequences were generated using different numbers of fireflies for a comparative analysis as demonstrated in Table 11.

| Fireflies=10 | | Fireflies=15 | | Fireflies=20 | |
|---|---|---|---|---|---|
| Path Priority | Cost | Path Priority | Cost | Path Priority | Cost |
| **19.6101** | **10.012** | **19.3830** | **10.0057** | **19.2750** | **10.0139** |
| **16.9001** | **10.025** | **16.7330** | **10.0311** | **16.6401** | **10.0144** |
| **8.3601** | **10.0188** | **8.3330** | **10.0005** | **8.3201** | **10.0225** |
| 4.7810 | 10.0751 | **4.7730** | 10.1094 | 4.7701 | 10.0851 |
| **13.867** | **10.0063** | **13.8012** | **10.0211** | **13.7580** | **10.0176** |
| 4.4501 | 10.0530 | 4.4330 | 10.1195 | 4.4380 | 10.0065 |
| 14.6817 | **10.0435** | 11.30531 | **10.05755** | 8.94933 | **10.02667** |

**TABLE 8 ATM-One Transaction Objective and average values with 10, 15, and 20 fireflies deployed using MOFA.** Non dominated values are highlighted in bold. Test sequences are evaluated on a multi-objective condition, encompassing both path priority value and cost. Firefly value 10 has the best results, but are less superior that the firefly value of 5.

As a part of the MOFA experimentation, 10 fireflies were applied to the ATM transaction dataset and its state transitions to evaluate the coverage of the software tests. As shown in Figure 8, MOFA provided positive results for the multicriteria problem, including Path Priority and cost parameters. At the highest path priority of 19.60, MOFA provided an optimal cost of 10.012, slightly higher than the global minimum cost for the ATM transaction dataset. These results demonstrate the efficiency and effectiveness of the MOFA algorithm in maximizing path priority while reducing cost. simultaneously.

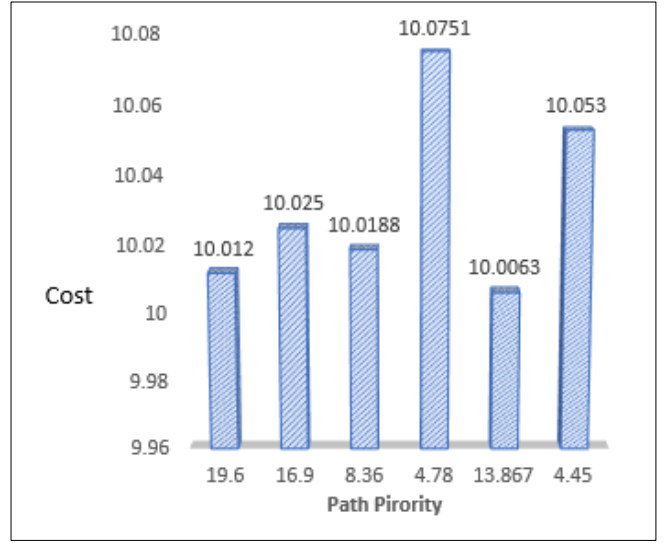

**FIGURE 8 Graphical representation of optimal test sequence generated while employing 10 fireflies with MOFA.** The highest path priority, at 19.6, demonstrated a reasonably lower (10.012) cost, fulfilling the objective of an effective software testing.

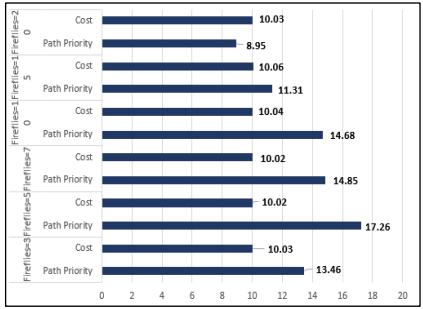

**FIGURE 7 Objective values for Multi-Objective Firefly algorithm.** Objective values including Oracle Cost and Path Priority for all non-dominated optimal test sequence for all particle values deployed using MOFA. FF number of 5 demonstrates the best ratio between both objectives, with the highest path priority of 17.26 and the lowest cost at 10.02.

| Fireflies=3 | | Fireflies=5 | | Fireflies=7 | |
|---|---|---|---|---|---|
| Path Priority | Cost | Path Priority | Cost | Path Priority | Cost |
| **21.250** | **10.024** | **20.251** | **10.018** | **19.893** | **10.017** |
| **18.200** | **10.006** | 17.440 | 10.027 | 17.143 | 10.021 |
| 8.600 | 10.020 | 8.482 | 10.025 | **8.401** | **10.019** |
| 4.867 | **10.069** | 4.841 | 10.009 | 4.801 | 10.084 |
| **14.389** | **10.034** | 14.101 | **10.018** | **13.977** | **10.032** |
| 4.500 | 10.073 | 4.500 | 10.015 | 4.464 | 10.063 |
| 13.461 | **10.031** | 17.263 | **10.021** | 14.853 | **10.022** |

**TABLE 5 ATM-One Transaction average Objective values with 3, 5, and 7 fireflies deployed using MOFA.** Non dominated values are highlighted in bold. Test sequences are evaluated on a multi-objective condition, encompassing both Path Priority value and cost. Best set of values were attained on a firefly value of 5, demonstrating the optimum conditionality of high path priority and low cost for test sequence generation.

| Optimal Test Sequences Generated | Independent paths generated | Objective | |
|---|---|---|---|
| | | Path Priority | Cost |
| **1,2,7,8** | 1,2,7,8 | **19.6001** | **10.0120** |
| 2,3,7 | 1,2,3,7,8 | 16.9001 | 10.0250 |
| 3,5,7 | 1,2,3,5,7 | 8.3602 | 10.0188 |
| 3,4,7 | 1,2,3,4,7 | 4.7801 | 10.0751 |
| **3,5,6,7** | 1,2,3,5,6,7 | **13.867** | **10.0063** |
| 4,8 | 1,2,4,8 | 4.4501 | 10.0530 |
| Average Values | | 12.1910 | 10.04355 |

**TABLE 10. ATM-One Transaction Objective values with Fireflies=10 using MOFA.** Non-dominated test sequences are highlighted in bold. Average values for multi-objective i.e., Path Priority and Oracle Cost, have been calculated only with the non-dominated values since they present non-redundant test sequences and hence are more reliable for software testing.

### E. RESULTS OF ATM-ONE TRANSACTION STATE TRANSITION

As the experiments show, PSO yields only one test sequence (3rd), with the highest path priority (0.8562) among all paths with any number of particles. Employing FA yields only one test sequence (1st) that has the highest path priority (19.275) among all paths with any number of fireflies. In MOPSO, however, a test sequence's optimal path priority (0.3901) leads to the optimal cost (0.2520) for the solution. Multiple non-dominant solutions (test sequences) were found that provide the highest path priority at the minimum cost compared to the results of the MOFA technique. MOPSO provides the optimal test sequence with the highest priority and the lowest cost, which is lower than the cost incurred by the Multi-Objective Firefly algorithm. Most importantly, the conclusion from the experimental results highlights that MOPSO is less computationally intensive than MOFA in generating test sequences. In Tables 12, 13, and 14, the bold values represent the objective values of all the non-dominant solutions of the optimal test sequence generated by MOPSO and MOFA. The average of the non-dominant solutions, obtained by calculating the bold values, is given at the end of each column. Most importantly, these algorithms were computed on the ATM-One Transition dataset with multiple states. The optimal test sequences exhibited that several test cases were generated, which contained a different number of non-dominated optimal test sequences for each state transition diagram. Finally, all the generated results were compared with the particle swarm optimization and multi-objective particle swarm optimization, which were compared with the Firefly algorithm, and the multicriteria Firefly variant. The obtained results demonstrate that the proposed solution reduces the generation of redundant test sequences and identifies the paths necessary for comprehensive software testing. Moreover, experimental results have also shown that the proposed approach achieves better performance than the methods used in studies [21] and [29]. Most importantly, the experimental results show that MOPSO was executed in a shorter time than MOFA. Moreover, the results of PSO and FA considered only a single path priority objective, while MOPSO and MOFA considered both the cost and prioritization of the important paths in the state diagram.

| Path Priority and Cost | | | | | | | |
|---|---|---|---|---|---|---|---|
| MOPSO | | MOFA | | MOPSO | | MOFA | |
| Particles=3 | | Fireflies=3 | | Particles=5 | | Fireflies=5 | |
| Path Priority | Cost | Path Priority | Cost | Path Priority | Cost | Path Priority | Cost |
| **0.1001** | **2.4460** | 21.250 | 10.024 | **0.1500** | **0.4158** | 20.250 | 10.018 |
| **0.1003** | **0.1932** | 18.200 | 10.006 | **0.170** | **0.597** | 17.440 | 10.027 |
| **0.127** | **2.802** | 8.600 | 10.020 | 0.193 | 2.074 | 8.480 | 10.025 |
| **0.100** | **2.224** | 4.867 | 10.069 | 0.187 | 0.411 | 4.840 | 10.009 |
| **0.100** | **0.589** | 14.389 | 10.034 | **0.185** | **0.241** | 14.101 | 10.018 |
| 0.087 | 2.866 | 4.500 | 10.073 | 0.150 | 2.212 | 4.500 | 10.015 |
| **0.105** | **1.651** | 13.461 | **10.031** | 0.168 | 0.418 | 17.263 | **10.021** |

**TABLE 11** Comparative analysis between MOPSO and MOFA on ATM-One dataset with particle and firefly values = 3 and 5. Values highlighted in bold refer to non-dominated solutions. Analysis was conducted on two objectives: Path Priority and Oracle Cost. Both the algorithms demonstrated higher path priority values and a lower cost simultaneously for higher particle and firefly numbers.

| Path Priority and Cost | | | | | | | |
|---|---|---|---|---|---|---|---|
| MOPSO | | MOFA | | MOPSO | | MOFA | |
| Particles=7 | | Fireflies=7 | | Particles=10 | | Fireflies=10 | |
| Path Priority | Cost | Path Priority | Cost | Path Priority | Cost | Path Priority | Cost |
| **0.237** | **0.869** | 19.893 | 10.017 | 0.362 | 1.069 | 19.601 | 10.012 |
| **0.260** | **0.886** | 17.143 | 10.021 | **0.390** | **0.252** | 16.901 | 10.025 |
| **0.290** | **2.040** | 8.401 | 10.019 | 0.431 | 1.282 | 8.360 | 10.018 |
| **0.285** | **0.371** | 4.801 | 10.084 | **0.425** | **0.224** | 4.780 | 10.075 |
| **0.278** | **1.750** | 13.976 | 10.032 | 0.414 | 1.297 | 13.867 | 10.006 |
| 0.250 | 1.266 | 4.464 | 10.063 | **0.362** | **0.351** | 4.450 | 10.053 |
| 0.266 | 1.386 | 14.853 | **10.022** | 0.392 | 0.591 | 14.681 | **10.043** |

**TABLE 12** Comparative analysis between MOPSO and MOFA on ATM-One dataset with particle and firefly values = 7 and 10. Values highlighted in bold refer to non-dominated solutions. Analysis was conducted on two objectives: Path Priority and Oracle Cost. MOPSO demonstrated higher path priority values and a lower cost simultaneously for higher particle numbers, contrary to MOFA which exhibited lower path priority and greater costs for higher firefly values.

| Path Priority and Cost | | | | | | | |
|---|---|---|---|---|---|---|---|
| MOPSO | | MOFA | | MOPSO | | MOFA | |
| Particles=15 | | Fireflies=15 | | Particles=20 | | Fireflies=20 | |
| Path Priority | Cost | Path Priority | Cost | Path Priority | Cost | Path Priority | Cost |
| **0.55010** | **0.61500** | 19.38301 | 10.0057 | 0.73750 | 0.60660 | 19.2750 | 10.0139 |
| **0.59010** | **0.33920** | 16.73310 | 10.0311 | 0.78000 | 0.32290 | 16.6401 | 10.0144 |
| **0.63750** | **0.18530** | 8.3331 | 10.0005 | 0.85620 | 0.48850 | 8.3201 | 10.0225 |
| **0.62750** | **0.21460** | 4.7730 | 10.1094 | 0.85000 | 0.61810 | 4.7701 | 10.0851 |
| **0.62140** | **0.05610** | 13.8010 | 10.0211 | 0.83570 | 0.68330 | 13.7580 | 10.0176 |
| 0.55000 | 0.81380 | 4.4330 | 10.1195 | 0.72500 | 0.59940 | 4.43801 | 10.0065 |
| **0.60532** | **0.28204** | 7.76283 | 10.05755 | 0.78707 | 0.47027 | 8.9493 | 10.02667 |

**TABLE 13** Comparative analysis between MOPSO and MOFA on ATM-One dataset with particle and firefly values = 15 and 20. Values highlighted in bold refer to non-dominated solutions. Analysis was conducted on two objectives: Path Priority and Oracle Cost. Both algorithms demonstrated higher path priority values for higher particle and firefly numbers. However, they also exhibited an unfavorable increase in the cost, rendering the respective firefly and particle values as unsuitable for test sequence generation.

## V. CONCLUSION

The proposed work focused on generating test sequences with multiple objectives while presenting a comparative analysis of meta-heuristic algorithms to demonstrate the superiority of algorithms with multiple objectives for the software testing problem. The dataset and state transition diagrams of ATM-One transaction were used for experimental purposes and evaluated with the Particle Swarm Optimization, the Firefly algorithm with a single objective of Path priority, the Multi-Objective Particle Swarm Optimization, and the Multi-Objective Firefly algorithm with Path Priority and Oracle Cost objectives. The results evidence that while the Particle Swarm Optimization provides an optimal test sequence for a particle count of 20, the Multi-Objective Optimization algorithm considers multiple objectives and produces test sequences that satisfy both the objectives at the same time. Multi-Objective Particle Swarm Optimization provides test sequences with higher priority and minimal execution cost than MOFA. It classifies the non-dominant solutions from a set of possible paths and generates non-redundant optimal test sequences with minimal computational cost while ensuring complete coverage of software testing procedures, thus providing an effective testing mechanism for software development.

**Author Contributions:** Conceptualization, Z.I., A.I., A.K. and K.Z.; methodology, Z.I., A.I., A.K. and K.Z.; software, Z.I., A.I., A.K. and K.Z.; validation, Z.I., A.I., A.K. and K.Z.; formal analysis, Z.I., A.I., A.K., M.T. and K.Z.; investigation, Z.I., A.I., A.K. and K.Z.; resources, Z.I., A.I., A.K. and K.Z.; data curation, Z.I., A.I., A.K. and K.Z.; writing—original draft preparation, Z.I., A.I., A.K. and K.Z.; writing—review and editing, Z.I., A.I., A.K. and K.Z.; visualization, Z.I., A.I., A.K. and K.Z.; supervision, K.Z. All authors have read and agreed to the published version of the manuscript.

**Funding:** This research received no external funding.

**Data Availability Statement:** Not applicable.

**Conflicts of Interest:** The authors declare no conflict of interest.